\documentstyle[12pt]{article}
\begin{document}
\begin{center}
\large
{\bf Types of Particle Oscillations  and Their
Realizations in $K^o$ and $\nu$ Oscillation}\\

Kh. M. Beshtoev\\

Joint Instit. for Nucl. Reser., Joliot Curie 6, 141980 Dubna,
Moscow region, Russia\\
\end{center}
\large
{\bf Abstract}
\par
Two particle vacuum transitions (oscillations) are studied in the general
case. We found that: 1) a nondiagonal mass term characterizing oscillations
is the width of two particle transitions into each other (this width
can be computed by the standard method); 2) two types of oscillations
take place: real and virtual.
\par
Solution of the problem of origin of mixing angle in the theory of
vacuum oscillations is given.
\par
It is shown that $K^{o}$-meson and neutrino oscillations
must proceed via two stages.
First, $K^{o}, \bar{K}^{o}$-eigenstates of strong interaction (or
$\nu_{e}, \nu_{\mu }, \nu_{\tau}$-eigenstates of weak interactions)
are  created.  Then, owing to the strangeness violating
weak interaction (or the lepton number violating interactions), these
meson states (or neutrino states) are converted into superpositions of
$K^{o}_1, {K}^{o}_2$-eigenstates of the weak interaction violating
strangeness (or $\nu_{1}, \nu_{2}, \nu_{3}$-eigenstates of the
interaction violating lepton numbers). Further, $K^o$-meson or neutrino
oscillations will  occur  in accordance with the standard scheme.


\section{Introduction}

In the old theory of neutrino oscillations [1, 2], constructed in
analogy with the theory of $K^{o}, \bar{K}^{o}$ oscillation, it
is supposed that mass eigenstates are $\nu_{1}, \nu_{2}, \nu_{3}$
neutrino states but not physical neutrino states
$\nu_{e}, \nu_{\mu }, \nu_{\tau}$, and that the neutrinos
$\nu_{e}, \nu_{\mu }, \nu_{\tau}$ are created as superpositions
of $\nu_{1}, \nu_{2}, \nu_{3}$  states.  This means that the
$\nu_{e}, \nu_{\mu }, \nu_{\tau}$ neutrinos have no definite mass, i.e.
their masses may vary depending on the  $\nu_{1}, \nu_{2}, \nu_{3}$
admixture  in the $\nu_{e}, \nu_{\mu }, \nu_{\tau}$  states
(naturally, in this case the low of conservation the energy and
the momentum of the neutrinos is not fulfilled). Besides, every particle
has mass shell and it will be left on its mass shell at passing through
vacuum.
Probably, this picture is incorrect one.
\par
In this work we consider particle mixings (oscillations).
At first we will consider mass matrix method for studding mixings
(oscillations) two particles in general case and then come to detailed
consideration of $K^o, \bar K^o$ and $\nu$ mixings (oscillations).

\section{Probability of $a \stackrel{B}\longrightarrow b$ Vacuum Transitions
(Oscillations)}
\par
Let us consider two particles (states) $a, b$ having numbers
(it can be $K^o, \bar K^o;
K^o_1, K^o_2$ or $\nu_e, \nu_\mu$) which can transit each into the other.
We can use the mass matrix of $a, b$ particles for consideration
of transitions
between these particles in the framework of the quantum theory (or particle
physics) since the mass matrix is eigenstate of a type of interaction
which creates these particles (see below).
\par
The mass matrix of $a$ and $b$ particles has the form
$$
\left(\begin{array}{cc} m_a & 0 \\ 0 & m_b  \end{array} \right) .
\eqno(1)
$$
\par
Due to the presence of a interaction violating their numbers,
a nondiagonal term appears in this matrix and then this mass
matrix is transformed in the following nondiagonal matrix
($CP$-is conserved):
$$
\left(\begin{array}{cc}m_a & m_{a b} \\ m_{a b} & m_b  \end{array}
\right)   ,
\eqno(2)
$$
which is diagonalized by turning through the angle $\beta$ and [2]
$$
tg 2\beta = \frac{2m_{a b}}{\mid m_a - m_b \mid}   ,
$$
$$
sin 2\beta = \frac{2m_{a b}}{\sqrt{(m_a - m_b)^2 +(2m_{a b})^2}}  .
\eqno(3)
$$
$$
\left(\begin{array}{cc} m_{a'} & 0 \\ 0 & m_{b'}  \end{array} \right) .
$$
$$
m_{a', b'} = {1\over 2} \left[ (m_{a} + m_{b})
\pm \left((m_{a} - m_{b})^2 + 4 m^{2}_{a b} \right)^{1/2} \right] ,
$$
\par
It is interesting to remark that expression (3) can be obtained from the
Breit-Wigner distribution [3]
$$
P \sim \frac{(\Gamma/2)^2}{(E - E_0)^2 + (\Gamma/2)^2}
\eqno(4)
$$
by using the following substitutions:
$$
E = m_b,\hspace{0.2cm} E_0 = m_a,\hspace{0.2cm} \Gamma/2 = 2m_{a b} ,
\eqno(5)
$$
where $\Gamma/2 \equiv W(... )$ is width of $a \rightarrow b$ transition,
then we can use standard method [4] for computing this value.
\par
We can see that here take place two cases of $a, b$
transitions (oscillations): real and virtual oscillations.
\par
1. If we consider the real transition of
$a$ into $b$ particle then
$$
sin^2 2\beta \cong \frac{4m^2_{a b}}{(m_a - m_b)^2 + 4m^2_{a b}}  ,
\eqno(6)
$$
if the probability of the real transition of
$a$ particles into $b$ particles through a interaction (i.e. $m_{a b}$) is
very small then
$$
sin^2 2\beta \cong \frac{4m^2_{a b}}{(m_a - m_b)^2} \cong 0 .
$$
\par
How can we understand this real $a \rightarrow b$ transition?
\par
If $2m_{a b} = \frac{\Gamma}{2}$ is not zero, then it means that
the mean mass of $a$ particle is $m_a$ and this
mass is distributed by $sin^2 2\beta$ (or by the Breit-Wigner formula) and
the probability of the $a \rightarrow b$ transition differs from zero
and it is defined by masses of $a$ and $b$ particles and $m_{a b}$, which
is computed in the framework of the standard method as it is pointed out
above.
\par
So, this is a solution of the problem of origin of mixing angle in the
theory of vacuum oscillations.
\par
In this case probability of $a \rightarrow b$ transition (oscillation)
is described by the following expression:
$$
P(a \rightarrow b, t) =  sin^2 2\beta sin^2
\left[\pi t\frac{\mid m_{b'}^2 - m_{a'}^2 \mid}{2 p_a} \right ] ,
\eqno(7)
$$
where $p_a$ is momentum of $a$ particle.
\par
2. If we consider the virtual transition of $a$ into $b$ particle then, since
$m_a = m_b$,
$$
tg 2\beta = \infty  ,
$$
i.e. $\beta = \pi/4$, then
$$
sin^2 2\beta = 1     .
\eqno(8)
$$
\par
In this case probability of $a \rightarrow b$ transition (oscillation) is
described by the following expression:
$$
P(a \rightarrow b, t) =
\left[\pi t\frac{4 m_{a b}^2}{2 p_a} \right ] ,
\eqno(9)
$$
\par
To make these virtual oscillations real their
participant in quasielastic interactions is necessary
for their transitions to own mass shells [5].
\par
It is clear that the process of $a \rightarrow b$ transition is a dynamical
process and at the beginning (i.e. at t = 0) here is no superposition
of $a', b'$
particles (states).
\par
Let us pass to consideration of concrete transitions (oscillations) between
different type particles (states).

\section{ $K^{o}, \bar{K}^{o}$- oscillations}

\par
1) The $K^{o},
\bar{K}^{o}$-mesons, which consist of  the $s, \bar{s}, d, \bar{d}$
quarks, are created in the strong interactions (the typical time of strong
interactions are $t_{\hbox{str}}\cong  10^{-23}$  s.  ) and are,
accordingly, eigenstates of these interactions, i.e. the mass matrix of
the $K^{o}, \bar{K}^{o}$ mesons is diagonal.
\par
2) If we take into
account the weak interaction (typical times of weak interactions are
$t_{\hbox{weak}} \cong  10^{-8}$  s.)  which  violates strangeness, then the
mass matrix of $K^{o}$-mesons will   become nondiagonal. If we
diagonalize this  matrix, then we will come to the $K^{o}_{1}, K^{o}_{2}$
states, which are eigenstates of the weak interaction [1].
\par
So we can see that, if $K^{0}$-mesons  are created in strong
interactions, then $K^{o}, \bar{K}^{o}$ mesons are produced, and if $K^{o}$
mesons are created in weak interactions then $K^{o}_{1}, K^{o}_{2}$  mesons
are created. In second case no  oscillations of $K^{o}$ mesons will
occur.
\par
Now let us  to give a phenomenological description of
$K^{o}, \bar{K}^{o}$ meson creation and oscillation  processes.  We
will consider the creation of $K^{o}, \bar{K}^{o}$-mesons as  a
quasistationary  process  with a typical time $t_{\hbox{str}}$. Within
of  this  typical  time $- t_{\hbox{str}}$, weak interactions will
violate strangeness and result in the mass matrix of the $K^{o}$-mesons
becoming nondiagonal. The probability for this process to occur in $t =
\pi  t_{\hbox{str}}$ is:
\par
$$
W(t = \pi \Delta t_{str}) = \frac{(1
- e^{-{t\over \Delta t_{str}})}} {(1 - e^{-{t\over \Delta t_{str}})}}
 \simeq  \pi  {\Delta t_{str} \over \Delta t_{weak}}
 \simeq  \pi  \cdot 10^{-15} ,
\eqno(10)
$$
\noindent
where $( 1- \exp ( - {t\over t_{str, weak}}))$
- is the decay probability of the quasistationary state during the time
$-t$.
\par
The  mass  matrix  of the $K^{o}$-mesons will become nondiagonal in
$t = \pi  10^{-23}$ s.  with a probability  of $W \cong  \pi  10^{-15}$.
And then the $K^{o}_{1}, K^{o}_{2}$  mesons--eigenstates of weak
interactions will  be  created.  So we can see that in this case mainly
$K^{o}, \bar{K}^{o}$ mesons will be produced but not the
$K^{o}_{1}, K^{o}_{2}$-mesons.
\par
3) Then, when the $K^{o}, \bar{K}^{o}$ mesons, that were
created  in  strong interactions,  pass  through  vacuum, the  mass  matrix
of  the $K^{o}$  mesons  will   become  nondiagonal, owing to
the presence  of  weak interactions  violating  strangeness. Diagonalizing
it, we   get $K^{o}_{1}, K^{o}_{2}$-meson  states  which are eigenstates
of weak interactions.  Obviously, the $K^{o}, \bar{K}^{o}$ mesons are, then,
 converted in to superpositions of $K^{o}_{1}, K^{o}_{2}$ -mesons
\par
$$
K^{o} = {{K^o_1 + K^o_2}\over \sqrt{2}}, \qquad
\bar K^{o} = {{K^o_1 - K^o_2}\over \sqrt{2}} .
\eqno(11)
$$
\noindent
Then, oscillations  of  the $K^{o}, \bar{K}^{o}$ mesons will  take
place  on a background of $K^{o}_{1}, K^{o}_{2}$  decays.
The length of these oscillations  is [1, 6]:
\par
$$
L_{osc} = {{2.48 p_{K^o}(MeV)}\over \mid m_{K^o_1} -
m_{K^o_2} \mid^2 (eV)^2}
\eqno(12)
$$
\noindent
$p_{K}o$  is the  momentum of $K^{o}$.
\par
The  main  question  which  arises now  is:  which   type   of
oscillations  real  (implying  actual  transitions    between   the
particles)  or  virtual  (implying  virtual   transitions   between
particles without transition to mass shells) take place between the
$K^{o}, \bar{K}^{o}$-mesons? Since the masses of $K^{o}$ and
$\bar{K}^{o}$  mesons  are  equal,
oscillations between these mesons are real. But, if the  masses  of
$K^{o}$ and $\bar{K}^{o}$ mesons were not equal, then the  oscillations
would  be virtual (the case of $K^o_1, K^o_2$ transitions was considered
in [7]).
\par
So, the mixings (oscillations) appear since at creating of $K^o$ mesons
are realized eigenstates of the strong interaction (i.e. $K^o, \bar
K^o$ mesons) but not eigenstates of the weak interaction violating
strangeness (i.e. $K^o_1, K^o_2$ mesons) and then, when they pass
through vacuum they are converted into superpositions of $K^o_1, K^o_2$
mesons. If $K^o_1, K^o_2$ mesons were originally created then mixings
(oscillations) would not take place since the strong interaction conserves
strangeness and isospin.

\section{ $\nu $-oscillations}
\par
We   can   now   pass   to   the   analysis    of three  neutrino
oscillations, taking  advantage  of  the  example  of
$K^{o}, \bar{K}^{o}$-meson oscillations.
\par
1)  The  physical  states  of  the $\nu_{e}, \nu_{\mu }, \nu_{\tau}$
neutrinos   are eigenstates of the weak interaction and, naturally,
the mass
matrix of $\nu_{e}, \nu_{\mu }, \nu_{\tau}$ neutrinos is diagonal.
All  the  available, experimental results indicate  that  the  lepton
numbers $l_{e}, l_{\mu }, l_{\tau}$  are   well conserved i.e. the standard
weak interactions do  not  violate  the lepton numbers.
\par
2) Then, to violate the  lepton  numbers, it  is  necessary  to introduce an
interaction violating these numbers. It is  equivalent to introducing
nondiagonal  mass terms  in the  mass  matrix  of $\nu_{e}, \nu_{\mu },
\nu_{\tau}$. Diagonalizing this matrix we go to the $\nu_{1}, \nu _{2},
\nu_{3}$ neutrino states. Exactly like the case  of $K^{o}$  mesons
creating  in strong interactions, when mainly $K^{o}, \bar{K}^{o}$  mesons
are produced, in  the considered case $\nu_{e}, \nu_{\mu }, \nu_{\tau}$,
but not $\nu_{1}, \nu_{2}, \nu_{3}$, neutrino  states  are mainly created
in the weak interactions (this is so, because   the contribution of the
lepton numbers violating interactions  in this process is too small).
And in the case 2) no oscillations take place.
\par
3) Then, when the $\nu_{e}, \nu_{\mu }, \nu_{\tau}$  neutrinos   pass
through vacuum, they  will  be  converted  into superpositions  of  the
$\nu_{1}, \nu _{2}, \nu_{3}$  owing  to presence  of  the
interactions violating  the  lepton number of neutrinos and  will be left
on  their mass   shells.  And, then, oscillations of the
$\nu_{e}, \nu_{\mu}, \nu_{\tau}$ neutrinos will  take  place according to
the standard scheme [1]. Whether these oscillations are real or virtual will
be determined by the masses of the  physical neutrinos
$\nu_{e}, \nu_{\mu}, \nu_{\tau}$.
\par
i) If the masses of the $\nu_{e}, \nu_{\mu }, \nu_{\tau}$ neutrinos  are
equal, then real oscillation of the neutrinos will take  place.
\par
ii) If  the masses  of  the $\nu_{e}, \nu _{\mu }, \nu _{\tau}$ are  not
equal, then virtual oscillation of  the  neutrinos will  take place.
To make these oscillations  real,  these  neutrinos must participate  in the
quasielastic interactions, in order to undergo transition  to  the mass shell
of the other appropriate neutrinos by analogue with $\gamma  - \rho ^{o}$
transition  in the  vector   meson  dominance model.  In case  ii)
enhancement of neutrino oscillations will  take place  if  the neutrinos pass
through  a bulk of matter [8].
\par
So, the mixings (oscillations) appear since at neutrinos creating
are realized eigenstates of the weak interaction
(i.e. $\nu_e, \nu_\mu, \nu_\tau$ neutrinos)
but not eigenstates of the weak interaction violating
lepton numbers (i.e. (i.e. $\nu_1, \nu_2, \nu_3$ neutrinos) and then,
when they pass through vacuum they are converted into superpositions of
$\nu_1, \nu_2, \nu_3$ neutrinos.
If $\nu_1, \nu_2, \nu_3$ neutrinos were originally created then mixings
(oscillations) would not take place since the weak interaction conserves
lepton numbers.
\par
The above considered approach for consideration of mass mixings is
the mass mixings approach besides of this approach is an another approach,
the charge mixings one, which is used in the vector dominance model [9].

\section{Conclusion}

Two particle vacuum transitions (oscillations) were studied in the general
case. We found that: 1) a nondiagonal mass term characterizing oscillations
is the width of two particle transitions into each other (this width
can be computed by the standard method); 2) two types of oscillations
take place: real and virtual.
\par
Solution of the problem of origin of mixing angle in the theory of
vacuum oscillations was given.
\par
It was shown that $K^{o}$-meson and neutrino oscillations
must proceed via two stages.
First, $K^{o}, \bar{K}^{o}$-eigenstates of strong interaction (or
$\nu_{e}, \nu_{\mu }, \nu_{\tau}$-eigenstates of weak  interactions)
are  created.  Then, owing to the strangeness violating
weak interaction (or the lepton number violating interactions), these
meson states (or neutrino states) are converted into superpositions of
$K^{o}_1, {K}^{o}_2$-eigenstates of the weak interaction violating
strangeness (or $\nu_{1}, \nu_{2}, \nu_{3}$-eigenstates of the
interaction violating lepton numbers). Further, $K^o$-meson or neutrino
oscillations will  occur  in accordance with the standard scheme.
\par
If $K^{o}_1, {K}^{o}_2$ mesons (or $\nu_1, \nu_2, \nu_3$ neutrinos)
were originally created then mixings
(oscillations) would not take place since the strong interaction
(or the weak interaction) conserves strangeness and isospins
(or lepton numbers).
\par
{\bf References}
\par
\noindent
1. F. Boehm, P. Vogel, Cambridge Univ. Press, 1987, p.27, p.121.
\par
S.M.Bilenky, S.T.Petcov, Rev. of Mod.  Phys., 59(1997)631.
\par
\noindent
2. V. Gribov, B.M. Pontecorvo, Phys. Lett.B, 28(1969)490.
\par
\noindent
3. J.M. Blatt, V.F Waiscopff, The Theory of Nuclear Reactions,
\par
INR T.R. 42.
\par
\noindent
4. L. B.  Okun, Leptons and Quarks, M, 1990.
\par
\noindent
5. Kh. M. Beshtoev, JINR Commun., E2-98-387, Dubna, 1998.
\par
\noindent
6. Kh. M. Beshtoev, Hadronic Journ. 1995, 18(1995)165.
\par
\noindent
7. Kh. M. Beshtoev, JINR Commun., E2-92-318, Dubna, 1992.
\par
Chenese Journ. of Phys., 34(1996)979.
\par
\noindent
8. Kh.M. Beshtoev, IL Nuovo Cim., 108A(1995)275.
\par
\noindent
9. J.J. Sakurai, Currents and Mesons, The Univ. of Chicago, 1967.
\par
Kh. M. Beshtoev, INR AC USSR Preprint, P-217, Moscow,
\par
1981.

\end{document}